\documentclass[
reprint, 
superscriptaddress,
 amsmath,
 amssymb,
 aps,
prb,
longbibliography
]{revtex4-2}

\usepackage{placeins}
\usepackage[percent]{overpic}
\usepackage{cuted}

\usepackage{graphicx}
\graphicspath{ {./figures/} }

\usepackage{dcolumn} 
\usepackage{bm}
\usepackage[colorlinks=true,
            linkcolor=blue,
            citecolor=blue,
            urlcolor=magenta]{hyperref}

\usepackage[mathlines]{lineno} 

\begin{document}

\preprint{APS/123-QED} 

\title{Metallic crossover through the tilt-free transition in La$_3$Ni$_2$O$_7$ at high pressure and temperature}

\author{Bastien Michon}
\thanks{Corresponding author: \href{mailto:bastien.michon@univ-tours.fr}{bastien.michon@univ-tours.fr}}
\affiliation{SOLEIL Synchrotron, L'Orme des Merisiers, RD 128, Saint Aubin 91190, France}
\affiliation{Present address: GREMAN - UMR7347 CNRS, Université de Tours, INSA Centre Val de Loire, Parc de Grandmont, Tours 37200, France}

\author{Yingpeng Yu}
\affiliation{Beijing National Laboratory for Condensed Matter Physics, Institute of Physics, Chinese Academy of Sciences, Beijing 100190, China}
\affiliation{School of Physical Sciences, University of Chinese Academy of Sciences, Beijing 100190, China}

\author{Beatrice D'Alò}
\affiliation{SOLEIL Synchrotron, L'Orme des Merisiers, RD 128, Saint Aubin 91190, France}

\author{Elena Stellino}
\affiliation{Department of Basic and Applied Sciences for Engineering, Sapienza University of Rome, Rome 00185, Italy}

\author{Gergely N\'emeth}
\affiliation{SOLEIL Synchrotron, L'Orme des Merisiers, RD 128, Saint Aubin 91190, France}

\author{Bosen Wang}
\affiliation{Beijing National Laboratory for Condensed Matter Physics, Institute of Physics, Chinese Academy of Sciences, Beijing 100190, China}
\affiliation{School of Physical Sciences, University of Chinese Academy of Sciences, Beijing 100190, China}

\author{Jianping Sun}
\affiliation{Beijing National Laboratory for Condensed Matter Physics, Institute of Physics, Chinese Academy of Sciences, Beijing 100190, China}
\affiliation{School of Physical Sciences, University of Chinese Academy of Sciences, Beijing 100190, China}

\author{Jinguang Cheng}
\affiliation{Beijing National Laboratory for Condensed Matter Physics, Institute of Physics, Chinese Academy of Sciences, Beijing 100190, China}
\affiliation{School of Physical Sciences, University of Chinese Academy of Sciences, Beijing 100190, China}

\author{Paolo Postorino}
\affiliation{Department of Physics, Sapienza University of Rome, Rome 00185, Italy}

\author{Ferenc Borondics}
\affiliation{SOLEIL Synchrotron, L'Orme des Merisiers, RD 128, Saint Aubin 91190, France}

\author{Francesco Capitani}
\affiliation{SOLEIL Synchrotron, L'Orme des Merisiers, RD 128, Saint Aubin 91190, France}


\begin{abstract}
La$_3$Ni$_2$O$_7$, a bilayer nickelate with Ruddlesden-Popper structure, undergoes a pressure-induced structural transition from a tilted \textit{Amam} phase to an untilted \textit{Fmmm} (or \textit{I4/mmm}) phase near 10-15 GPa, concomitant with the emergence of high-T$_c$ superconductivity ($T_c \sim 80$~K). Despite intense interest, the phase boundaries and the impact of structural changes on the electronic properties remain unclear. Here, we combine high-pressure and high-temperature Raman and synchrotron-based infrared spectroscopies to map the structural and electronic evolutions. Raman measurements confirm the pressure-driven structural transition and reveal the emergence of Fano line shapes, indicating enhanced electron-phonon coupling. High-temperature data show analogous spectral signatures above 544~K, suggesting an upper temperature limit of the \textit{Amam} phase within the T-P phase diagram of this system. Infrared reflectivity measurements evidence a concomitant enhanced metallicity, with a tremendous two-order-of-magnitude increase in carrier density, marking a crossover from a weakly to highly metallic state. These results establish a unified picture of the structural transition and its strong coupling to the electronic properties.

\end{abstract}

\maketitle




\textit{Introduction} -- Superconductivity in nickelates has long been sought, particularly in the perovskite series \textit{R}NiO$_3$ (\textit{R} = rare-earth cation) \cite{Pup26}. These compounds exhibit a complex phase diagram, including structural, metal-insulator and antiferromagnetic transitions, which depend on temperature (T) and the tolerance factor (\textit{t}) controlled by the size of the \textit{R} cation \cite{Cata18,Ardi21}.

Although superconductivity has not been observed in the perovskite structure, structural distortions -- particularly the tilting of NiO$_6$ octahedra -- are believed to prevent its emergence. This has led to the exploration of layered nickelates, where reduced dimensionality and altered NiO$_6$ connectivity can mitigate such distortions. In 2019, superconductivity was discovered in thin films of hole-doped infinite-layer \textit{R}NiO$_2$, with a T$_c \sim$ 15~K \cite{Li19,Osada21}.

Recently, superconductivity has also been reported in the bilayer Ruddlesden-Popper compound La$_3$Ni$_2$O$_7$ under pressure, with T$_c$ reaching up to 80~K at 10-15~GPa \cite{Sun23,GWang24}. At ambient conditions, the system crystallizes in a tilted orthorhombic \textit{Amam} structure, whereas pressure induces a structural transition that suppresses the tilts in NiO$_6$ octahedra, resulting in high-T$_c$ superconductivity \cite{Sun23}.

Understanding the relationship between lattice structure, electronic properties and superconductivity is crucial to unravel the mechanism of high-T$_c$ superconductivity in these systems. However, the high-pressure structure of La$_3$Ni$_2$O$_7$ remains debated, with conflicting reports suggesting either an orthorhombic \textit{Fmmm} or a tetragonal \textit{I4/mmm} phase based on X-ray diffraction and theoretical calculations \cite{GWang24,LWang24,GWang25,JLi25,Geisler24}. Recent Raman studies propose a continuous evolution under pressure, from a coexistence of \textit{Amam} and \textit{Fmmm} orthorhombic phases to a pure \textit{I4/mmm} tetragonal phase \cite{HZhang25}. As a result, the onset of this transition in the temperature-pressure (T-P) phase diagram remains incomplete \cite{Sun23,LWang24}. These structural changes are also expected to significantly influence free carriers, similar to perovskites where lattice distortions govern metal-insulator transitions \cite{Cata18}.

Here, we address these issues by performing high-pressure (HP) and high-temperature (HT) Raman spectroscopy on La$_3$Ni$_2$O$_7$ single crystals, complemented by infrared and visible reflectivity measurements, revealing pressure- and temperature-induced metallic crossover through the structural transition. Details on the crystal growth and the experimental setups are provided in the Supplemental Material~\cite{suppmat} (see also Refs.~\cite{FLi26supp,AKuz05supp} therein).

\begin{figure}[h]
\includegraphics[scale=0.24]{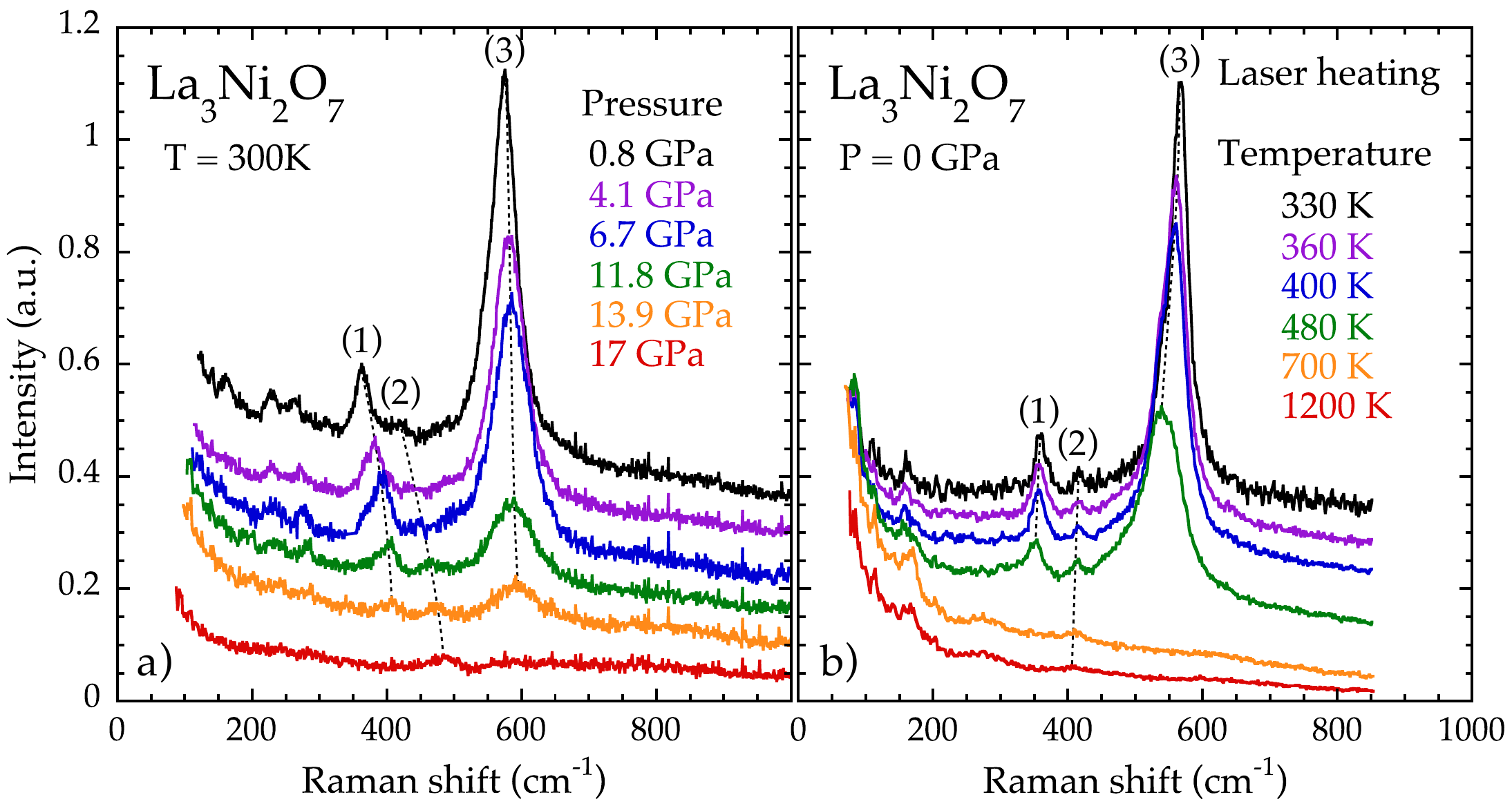}
\caption{\label{fig:fig1} Structural transition observed in Raman spectra under high pressure and high temperature (laser heating). Temperatures, derived from anti-Stokes thermometry (Methods~\cite{suppmat}), are rounded and correspond to the midpoints of their uncertainty ranges (see Fig.~\ref{fig:fig2}).}
\end{figure}

\textit{Structural Transition} -- We performed high-pressure Raman measurements up to 17~GPa in a diamond anvil cell at room temperature (T = 300~K), using a 532~nm green laser to probe the \textit{ab} plane of La$_3$Ni$_2$O$_7$. Spectra were collected in the 100-1000~cm$^{-1}$ range. Fig.~\ref{fig:fig1}a shows the Raman data at selected applied pressures (for further spectra see Fig.~S1~\cite{suppmat}). The Raman spectrum at 0.8~GPa (black curve) shows peaks ordered from low to high Raman shift: modes at 163~cm$^{-1}$, 233~cm$^{-1}$, and 360~cm$^{-1}$ (labeled (1)), followed by peaks at 419~cm$^{-1}$ (mode (2)) and 565~cm$^{-1}$ (mode (3)), the latter being the most intense. We observe three pressure-induced trends, all of which reverse upon pressure release:
\begin{itemize}
  \item All peaks gradually blueshift (black dotted lines).
  \item Main peaks (1) and (3) gradually broaden, disappearing completely around 14-15~GPa.
  \item Peak (1) develops an asymmetric Fano line shape above 5-6~GPa (see blue and green curves) \cite{UFano61}, deviating from a symmetric Lorentzian to a skewed profile (sawtooth-like in our case).
\end{itemize}

The appearance and gradual development of a Fano-like line shape in the Raman spectrum reflect the interference between the phonon excitation and an underlying electronic continuum \cite{UFano61,TDeve07}. This asymmetry, characteristic of electron-phonon coupling, becomes more pronounced above $\sim$~6~GPa. As pressure increases, the intensities of the peaks decrease, ultimately leading to the complete suppression of the Raman signal around 15~GPa. This marks the structural transition from the orthorhombic \textit{Amam} phase to a high-symmetry tilt-free phase.

By systematically varying the incident laser power, we observe a strong dependence of the Raman spectra on heating. Measurements at ambient pressure reveal spectral features similar to those obtained under high pressure (Fig.~\ref{fig:fig1}b), except that the Raman modes exhibit a redshift rather than a blueshift, consistent with lattice expansion induced by laser heating. The corresponding temperatures are determined from the Stokes/anti-Stokes intensity ratio (see Methods~\cite{suppmat}).

The temperature dependence of the Raman spectra reveals that the \textit{Amam} phase vanishes between 480 and 700~K, marking an upper-limit temperature for the structural transition in the T-P phase diagram of bilayer nickelates. This effect is fully reversible upon reducing laser power. Previous studies on nickelate perovskites (\textit{R}NiO$_3$) have shown that laser heating induces structural, metal-insulator and antiferromagnetic transitions \cite{Ardi21}. To quantify the observed effect, we conducted Raman spectroscopy on a La$_3$Ni$_2$O$_7$ sample mounted on a heating stage, which enabled precise determination of the onset temperature of the structural transition at ambient pressure to be 544~K (see Fig.~S2a,b~\cite{suppmat}), in strong agreement with a reported transition of 550~K observed decades ago in Xray diffraction, transport and magnetic susceptibility \cite{Koba96,Sasa97,Zink07}.

\begin{figure}[h]
\includegraphics[scale=0.25]{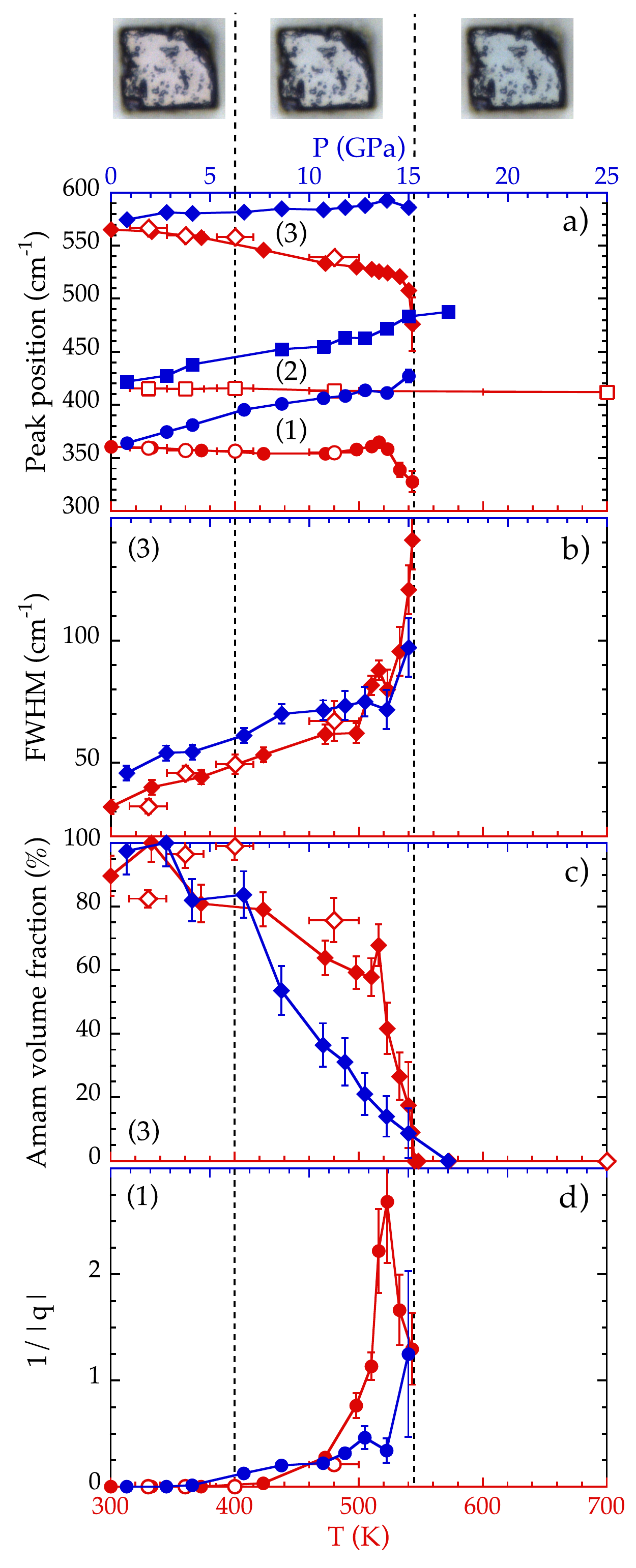}
\caption{\label{fig:fig2} Pressure and temperature evolutions of fitting parameters from Raman modes (1), (2) and (3). HP data are shown as blue symbols, HT data as red symbols, with a double x-scale allowing for comparison of pressure and temperature effects on the modes. Peak parameters for (1), (2) and (3) are represented by circles, squares and diamonds, respectively. Empty red symbols correspond to data from laser heating experiments, with temperatures determined by anti-Stokes thermometry (Methods~\cite{suppmat}).}
\end{figure}

\begin{figure*}[htp]
\includegraphics[scale=0.3]{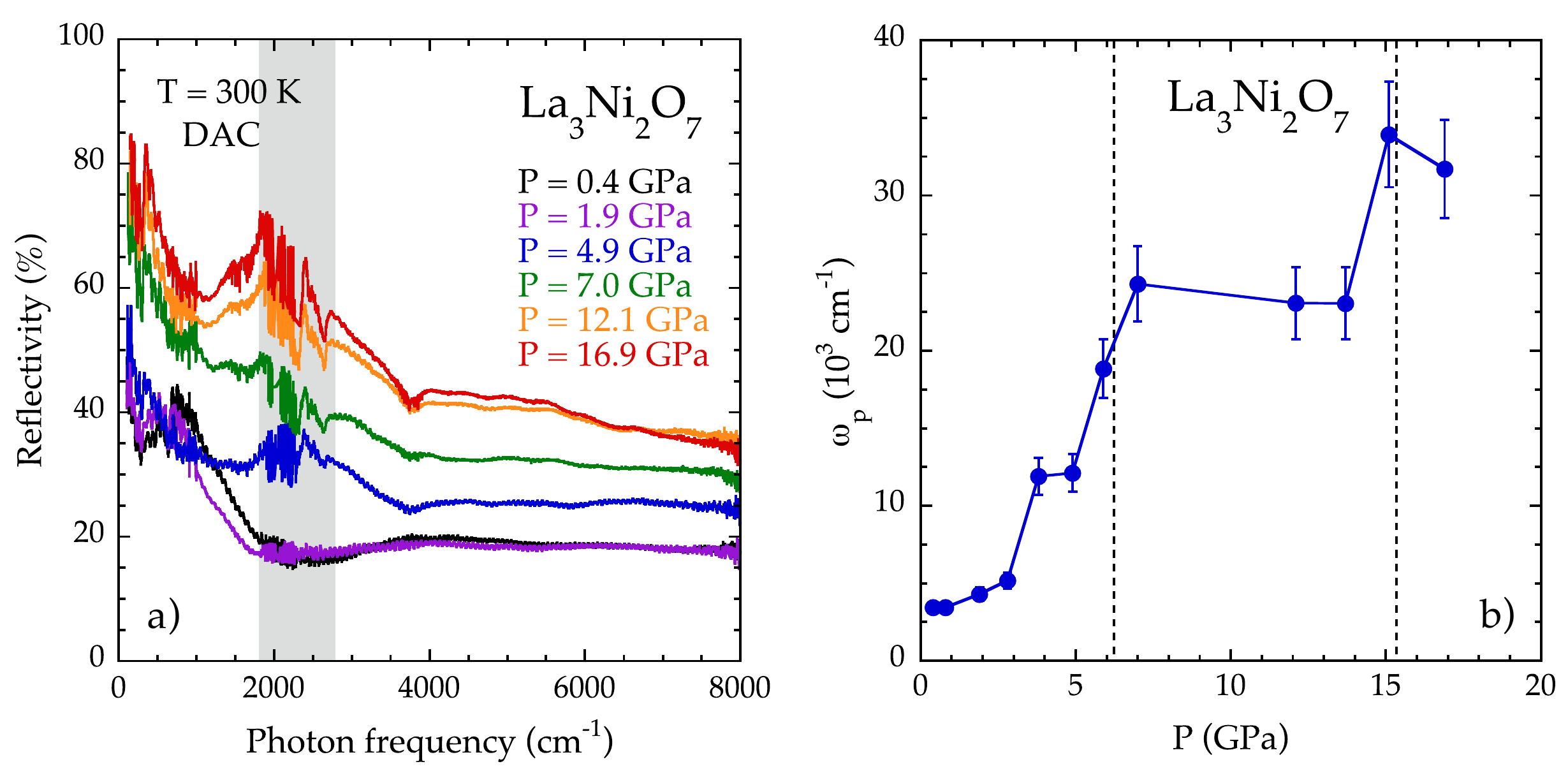}
\caption{\label{fig:fig3} Evidence for a metallic crossover from HP infrared reflectivity. The grayed region in panel (a) corresponds to strong absorptions by diamond phonons, which introduces significant noise and limits the reliability of the reference signal, particularly at HP (modifications in diamond phonon modes).}
\end{figure*}

In Fig.~\ref{fig:fig2}, we systematically analyze the HP-HT Raman dataset for modes (1), (2) and (3), tracking their peak positions, widths (FWHM), integrated intensities ($\propto$ peak size $\times$ FWHM) and Fano parameters $1/|q|$ under pressure and heating (see Figs.~S3, S4 and Methods~\cite{suppmat}). Phonon (2) remains stable up to 17~GPa and 1200~K without significant changes in width and intensity (see Fig.~S4~\cite{suppmat}). In contrast, phonons (1) and (3) broaden and vanish under pressure at 300~K and show similar behavior upon heating at 0~GPa:
\begin{itemize}
\item Phonons blueshift with pressure and redshift with increasing temperature up to 15.25~GPa/544~K, where they exhibit pronounced softening at the structural transition (Fig.~\ref{fig:fig2}a).
\item FWHM increases linearly with both pressure and temperature for phonon (3), and exhibits a dramatic damping near the transition (Fig.~\ref{fig:fig2}b), similarly for phonon (1) (Fig.~S3a~\cite{suppmat}).
\item Integrated phonon intensities remain constant at low pressure/temperature until 6~GPa/400~K, then progressively vanish at 15.25~GPa/544~K (Fig.~\ref{fig:fig2}c and Fig.~S3b~\cite{suppmat}).
\end{itemize}

When a phonon is associated with a specific structural phase, its relative integrated intensity serves as a proxy for the corresponding volume fraction \cite{Gyak19}. Phonon (3), originating from vibrations of tilted NiO$_6$ octahedra, acts as a fingerprint for the \textit{Amam} phase \cite{HZhang25}. By normalizing its integrated intensity to 100\% at low pressure and temperature, we track the evolution of the \textit{Amam} volume fraction (Fig.~\ref{fig:fig2}c; similarly with phonon (1) in Fig.~S3b~\cite{suppmat}). The data reveal that the sample remains fully in the \textit{Amam} phase below 6~GPa and 400~K. Above these conditions, a continuous transition occurs toward a tilt-free structure, with the \textit{Amam} phase disappearing completely above 15.25~GPa or 544~K. The intermediate region (6-15.25~GPa) corresponds to a coexistence regime of tilted \textit{Amam} and tilt-free phases. The coexistence clearly ends at 15.25~GPa (300~K) and 544~K (0~GPa), where phonons (1) and (3) exhibit strong softening and damping. The points (6~GPa, 300~K) and (0~GPa, 400~K) mark the onset of the coexistence region in the T-P phase diagram.

In the visible range, we observe a reproducible color change of the sample upon heating from 300~K to 673~K, shifting from pink to blue (see top images in Fig.~\ref{fig:fig2}). To rule out an artifact from the microscope camera, we performed visible reflectivity measurements. As shown in Fig.~S5~\cite{suppmat}, the reflectivity remains relatively unchanged between 300~K and 423~K, consistent with the pure \textit{Amam} phase. In contrast, a pronounced decrease in reflectivity occurs in the red (low-energy) region between 423~K and 573~K, with no further change above 573~K, \textit{i.e.}, beyond the structural transition. These results indicate a significant modification of the electronic structure, linked to changes in interband transitions across the transition.


\textit{Metallic Crossover} -- As discussed above, phonon (1) develops a Fano line shape upon increasing pressure or temperature. In Fig.~\ref{fig:fig2}d, we plot $1/|q|$, where $q$ is the Fano parameter (Methods~\cite{suppmat}), as a function of pressure (blue circles) and temperature (red circles). The quantity $1/|q|$ reflects the strength of the electron-phonon coupling. It remains negligible below 6~GPa and 400~K, and then increases progressively, reaching a maximum as the phonon disappears at 15.25~GPa or 544~K. For the temperature-dependent data, $1/|q|$ slightly decreases just below the transition, likely due to strong phonon softening and damping, which may affect the coupling strength. Previous studies have observed that some low-frequency phonons develop a Fano line shape and survive above the structural transition, with their asymmetry largely unchanged \cite{HZhang25}. This suggests that, starting from the coexistence region, the density of free carriers progressively increases, strengthening their coupling to phonons, before it saturates above the structural transition.

To investigate the pressure evolution of free charge carriers, we performed synchrotron infrared reflectivity measurements at 300~K in a diamond anvil cell up to 16.9~GPa (see Fig.~\ref{fig:fig3}a and further in Fig.~S6a~\cite{suppmat}). The reflectivity spectrum at 0.4 GPa (black curve) is consistent with previous reports, exhibiting a strong infrared-active phonon at 695~cm$^{-1}$ and a plasma edge at low frequency around 2500-3000~cm$^{-1}$ \cite{ZLiu24}, associated with the free carrier response. Upon increasing pressure, the phonon mode at 695~cm$^{-1}$ progressively disappears, while the reflectivity in the 2500-3000~cm$^{-1}$ range increases by approximately a factor of 3 at 16.9~GPa (red curve). Concomitantly, the plasma edge shifts to higher frequencies, beyond our measured spectral range.

We extracted the plasma frequency of free carriers, $\omega_p$, by performing a least-square fitting procedure of the reflectivity spectra, modeling the diamond interface with a dielectric constant $\varepsilon_d$ = 5.84 (refractive index $n_d = 2.42$) and the nickelate dielectric function with Drude-Lorentz contributions (see Methods~\cite{suppmat}). To constrain the high-frequency response, we combined room- and high-temperature visible reflectivity data (300~K and 673~K), assuming that HP and HT lead to similar reflectivity in the visible range. From the best-fit results (Fig.~S7~\cite{suppmat}), we obtain $\omega_p$ = 3450~cm$^{-1}$ at 0.4~GPa and $\omega_p \approx$ 32000~cm$^{-1}$ at 16.9~GPa. The pressure dependence of $\omega_p$ is shown with blue circles in Fig.~\ref{fig:fig3}b. Since $\omega_p^2 \propto n/m$, where $n$ is the carrier density and $m$ the effective mass, our analysis reveals an increase of nearly two orders of magnitude in free carrier density in the HP tilt-free phase compared to the \textit{Amam} phase, assuming a weak variation of the effective mass (see Fig.~S6b~\cite{suppmat}). Infrared spectroscopic measurements unambiguously demonstrate a strong increase in carrier density at the structural transition, consistent with the enhanced electron-phonon coupling reflected by the Fano parameter of the phonon modes, which highlights a metallic crossover from a low-carrier-density to a high-carrier-density metal.

\begin{figure}[htp]
\includegraphics[scale=0.3]{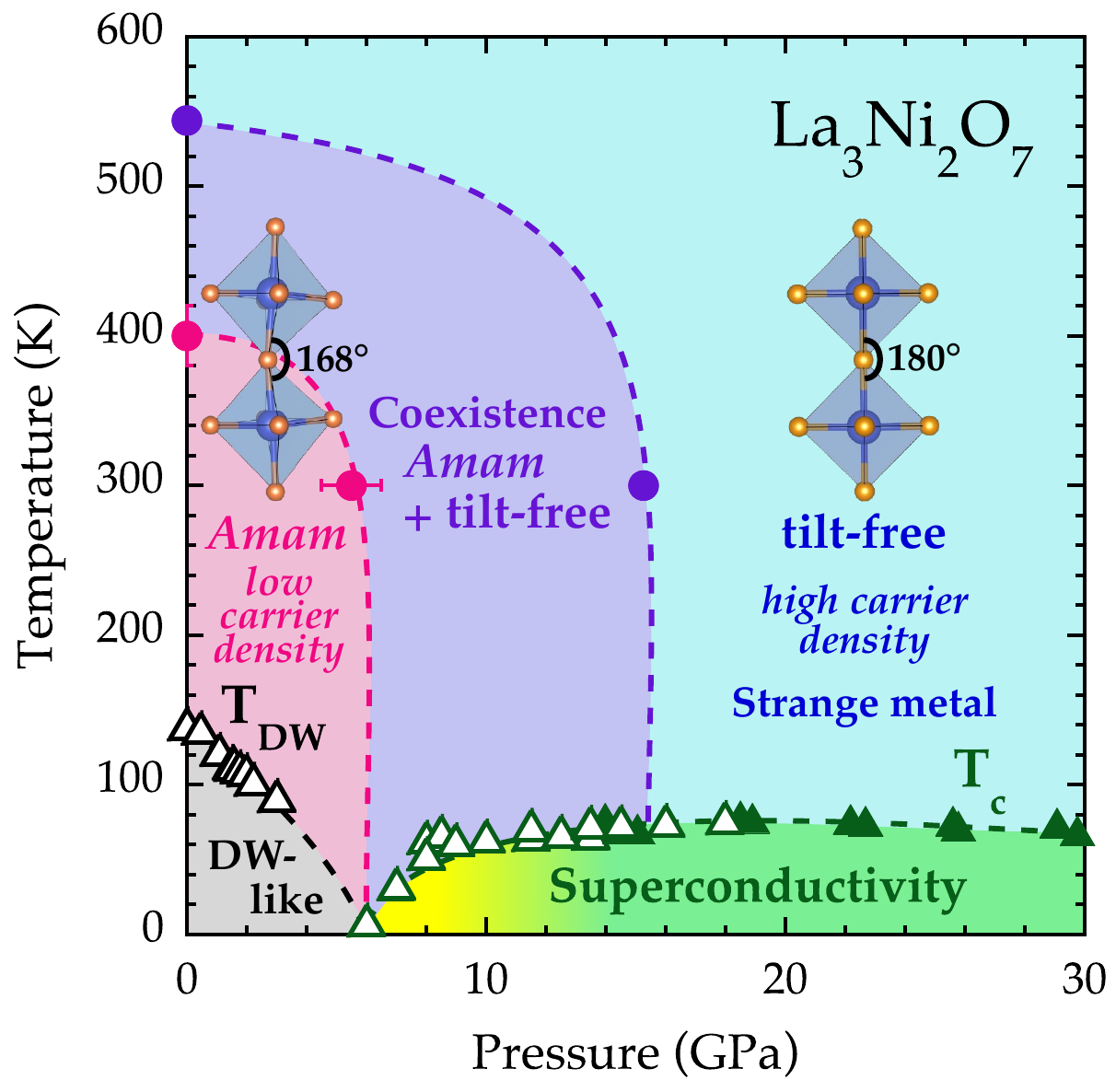}
\caption{\label{fig:fig4} T-P phase diagram constructed from pressure- and temperature-dependent Raman measurements. Black and green triangles indicate the onset temperatures of the DW-like order (T$_{DW}$) and superconductivity (T$_c$), respectively, with filled symbols for single crystals \cite{Sun23} and open symbols for polycrystals \cite{GWang24}. Pink and purple circles represent onset pressures and temperatures from our Raman data. The pink, purple and blue regions correspond to the \textit{Amam} (weakly metallic), coexistence (\textit{Amam} + tilt-free) and pure tilt-free (high-carrier-density metal) phases, respectively. Strange metallicity is also reported in the high-pressure region (blue) \cite{YZhang24}. A color gradient in the superconducting region highlights the crossover from filamentary to bulk superconductivity \cite{GWang24,NWang24}. Schematic representations of tilted and tilt-free NiO$_6$ octahedra are included (adapted from \cite{Sun23}).}
\end{figure}

\textit{Phase Diagram} -- We construct the T-P phase diagram (Fig.~\ref{fig:fig4}), identifying boundaries between the tilted \textit{Amam} phase, the coexistence region (\textit{Amam} + tilt-free) and the pure tilt-free phase, represented by the pink, purple and blue zones, respectively. The pink region corresponds to the tilted \textit{Amam} structure, associated with a weakly metallic state, while the blue one denotes the tilt-free structure, corresponding to a high-carrier-density metal. High-pressure and high-temperature Raman measurements define key points at (6~GPa, 300~K), (15.25~GPa, 300~K), (0~GPa, 400~K) and (0~GPa, 544~K), marked by pink and purple circles.

The dashed lines separating the regions serve as guides to the eye, connecting these points and suggesting vertical boundaries at low temperature, consistent with low-temperature X-ray diffraction measurements \cite{Sun23}. This phase diagram reveals a clear correspondence between the \textit{Amam} phase and the density-wave-like (DW-like) order at low pressure, and between the tilt-free phase and superconductivity at high pressure. Superconductivity emerges at the boundary of the coexistence region around 6~GPa, coinciding with an increase in carrier density. Previous studies suggest a crossover from filamentary superconductivity emerging at 6-7~GPa to bulk superconductivity near 15~GPa \cite{GWang24,NWang24}, consistent with the evolution of the structural transition. Superconductivity may initially develop within tilt-free domains embedded in the coexistence phase, leading to its filamentary aspect.

\textit{Summary} -- In this work, we combined pressure- and temperature-dependent Raman spectroscopy, complemented by high-pressure infrared reflectivity measurements, which together reveal a pressure- and temperature-driven transition from a tilted \textit{Amam} structure to a tilt-free phase.

Phonons at 360~cm$^{-1}$ (1) and 565~cm$^{-1}$ (3) disappear above 15.25~GPa or 544~K, accompanied by strong softening and damping (Fig.~\ref{fig:fig2}, Fig.~S3~\cite{suppmat}), providing clear evidence of the structural transition. The integrated intensity of these modes, tracking the \textit{Amam} volume fraction, remains nearly constant at low pressure/temperature, decreases within the coexistence region, and vanishes at the transition (Fig.~\ref{fig:fig2}c, Fig.~S3b~\cite{suppmat}). In parallel, the Fano parameter $1/|q|$ strongly increases above $\sim$~6~GPa or 400~K (Fig.~\ref{fig:fig2}d), together with the plasma frequency $\omega_p$ of the free electrons (Fig.~\ref{fig:fig3}a,b), highlighting a pressure- and temperature-induced metallic crossover.

These results establish a direct relation between the structural transition and the crossover from a weakly to highly metallic state with significantly enhanced carrier density. 

We also observe a close concomitance between the emergence of the tilt-free phase and superconductivity above $\sim$~6-7~GPa in the T-P phase diagram (Fig.~\ref{fig:fig4}), underlining the important role of the structure as a prerequisite for superconductivity. However, the stabilization of the tilt-free \textit{I4/mmm} phase at ambient conditions in high-pressure O$_2$-annealed samples does not induce superconductivity \cite{MShi25}, emphasizing that while high-symmetry structure and metallicity are important, they are insufficient by themselves to trigger superconductivity.

In layered nickelates, the presence of the DW-like order at low pressure, together with reports of linear-in-temperature resistivity, suggests the existence of strong electronic fluctuations extending into the highly metallic region, potentially contributing to a strange-metal behavior (Fig.~\ref{fig:fig4}) \cite{Sun23,YZhang24,Pup26}. This situation is reminiscent of cuprates, where infrared spectroscopy reveals a direct link between superconductivity and electronic correlations, together with signatures of Planckian dissipation in the strange-metal region \cite{BMichon21,BMichon23}.

Within this framework, fluctuations associated with the DW-like phase may provide a possible route to the superconducting pairing mechanism in the vicinity of the structural transition, although further work is required to establish their precise role. Overall, superconductivity in La$_3$Ni$_2$O$_7$ appears intimately connected to the structural-driven metallic enhancement, which defines the electronic environment in which charge pairing emerges.


\textit{Acknowledgments} -- B.M. coordinated the project. B.M. and F.C. defined the project objectives. B.M. and F.C. carried out the high-pressure Raman and infrared reflectivity measurements with assistance from B.D.. B.D., E.S. and P.P. performed Stokes and anti-Stokes Raman measurements with varying laser power. B.M., G.N. and F.B. conducted high-temperature Raman and visible reflectivity measurements. B.M. performed the analysis and carried out the interpretation of the entire dataset. F.C. and B.D. provided feedback on the analysis process. B.M. wrote the manuscript, incorporating comments from the coauthors. Y.Y., B.W., J.S. and J.C. synthesized and characterized the La$_3$Ni$_2$O$_7$ single crystals. F.B. provided postdoctoral funding for B.M.

The work at IOPCAS is supported by the National Key R\&D Program of China (Grant No. 2023YFA1406100), the National Natural Science Foundation of China (Grant Nos. 12522407, 12494592, and 12025408).

\FloatBarrier
\bibliography{biblio}

\end{document}


\preprint{APS/123-QED} 

\title{Supplementary: Metallic crossover through the tilt-free transition in La$_3$Ni$_2$O$_7$ at high pressure and temperature}

\author{Bastien Michon}
\thanks{Corresponding author: \href{mailto:bastien.michon@univ-tours.fr}{bastien.michon@univ-tours.fr}}
\affiliation{SOLEIL Synchrotron, L'Orme des Merisiers, RD 128, Saint Aubin 91190, France}
\affiliation{Present address: GREMAN - UMR7347 CNRS, Université de Tours, INSA Centre Val de Loire, Parc de Grandmont, Tours 37200, France}

\author{Yingpeng Yu}
\affiliation{Beijing National Laboratory for Condensed Matter Physics, Institute of Physics, Chinese Academy of Sciences, Beijing 100190, China}
\affiliation{School of Physical Sciences, University of Chinese Academy of Sciences, Beijing 100190, China}

\author{Beatrice D'Alò}
\affiliation{SOLEIL Synchrotron, L'Orme des Merisiers, RD 128, Saint Aubin 91190, France}

\author{Elena Stellino}
\affiliation{Department of Basic and Applied Sciences for Engineering, Sapienza University of Rome, Rome 00185, Italy}

\author{Gergely N\'emeth}
\affiliation{SOLEIL Synchrotron, L'Orme des Merisiers, RD 128, Saint Aubin 91190, France}

\author{Bosen Wang}
\affiliation{Beijing National Laboratory for Condensed Matter Physics, Institute of Physics, Chinese Academy of Sciences, Beijing 100190, China}
\affiliation{School of Physical Sciences, University of Chinese Academy of Sciences, Beijing 100190, China}

\author{Jianping Sun}
\affiliation{Beijing National Laboratory for Condensed Matter Physics, Institute of Physics, Chinese Academy of Sciences, Beijing 100190, China}
\affiliation{School of Physical Sciences, University of Chinese Academy of Sciences, Beijing 100190, China}

\author{Jinguang Cheng}
\affiliation{Beijing National Laboratory for Condensed Matter Physics, Institute of Physics, Chinese Academy of Sciences, Beijing 100190, China}
\affiliation{School of Physical Sciences, University of Chinese Academy of Sciences, Beijing 100190, China}

\author{Paolo Postorino}
\affiliation{Department of Physics, Sapienza University of Rome, Rome 00185, Italy}

\author{Ferenc Borondics}
\affiliation{SOLEIL Synchrotron, L'Orme des Merisiers, RD 128, Saint Aubin 91190, France}

\author{Francesco Capitani}
\affiliation{SOLEIL Synchrotron, L'Orme des Merisiers, RD 128, Saint Aubin 91190, France}

\maketitle


\clearpage

\section*{\label{sec:Metho}Section 1. Methods}

\subsection{Single-crystal growth}
Single crystals of La$_3$Ni$_2$O$_7$ were grown by a molten salt flux evaporating method at ambient pressure, which follows the procedure as described in Ref.~\cite{FLi26supp}. They are platelets with a square-shape (ab plane) of about 40-50~$\mu$m size and with a thickness around 10-30~$\mu$m (along the c-axis). High-purity La$_2$O$_3$ (99.99\%, dried at 1000~\textdegree{C} overnight) and NiO were used as starting materials. The oxide powders were weighed and grounded, and then mixed with anhydrous K$_2$CO$_3$ as flux in a solute-to-flux mass ratio of 1:15. To prevent moisture absorption, all the processes were performed inside a glovebox. The mixture was placed in an alumina crucible that was covered with a lid to control the evaporation rate. The crystals were grown in the furnace at a temperature of 1000-1050~\textdegree{C} for 72~h with evaporating the flux gradually and then were cooled to room temperature naturally. The crystals were extracted by soaking the mixture in deionized water. To reduce the oxygen vacancies and further improve the stoichiometry, we annealed the as-grown crystals in a tube furnace under a flowing oxygen atmosphere at 500~\textdegree{C} for 5 days, followed by furnace cooling to room temperature.

\subsection{Experimental setups}
For room-temperature high-pressure measurements, samples were loaded into a membrane diamond anvil cell (DAC) equipped with type-IIa diamonds with 600~$\mu$m culets. Stainless steel gaskets were pre-indented to $\sim$~50~$\mu$m, and a 250~$\mu$m hole was drilled by electrical discharge machining. The sample chamber was filled with CsI as pressure-transmitting medium, together with a $\sim$~50~$\mu$m nickelate crystal and a ruby sphere for \textit{in situ} pressure calibration via ruby luminescence.

High-pressure Raman and synchrotron infrared measurements were performed at the SMIS beamline of the SOLEIL synchrotron using a custom-built microscope coupled to a Fourier-transform infrared spectrometer (Thermo Scientific Nicolet iS50) and a Raman spectrometer (Horiba Jobin Yvon iHR320). Infrared spectra were acquired in reflectivity geometry using a KBr beamsplitter and a liquid-N$_2$-cooled MCT detector in the mid-infrared range, and a solid-substrate beamsplitter with a liquid-He-cooled bolometer in the far-infrared range. A gold foil placed between the anvils was used as the reference for reflectivity measurements.

The Raman spectrometer was fiber-coupled to a modified Horiba Super Head, using a 20$\times$ super-long-working-distance Mitutoyo objective and a 532~nm laser (Cobolt 08-DPL). Rayleigh scattering was rejected using a Semrock RazorEdge filter. Spectra were acquired with a 1800~lines/mm grating and a CCD detector, with an acquisition time of 300~s and two accumulations.

For ambient-pressure high-temperature measurements, the sample temperature was increased either by raising the incident laser power or by using an optical heating stage (Linkam) up to 400~\textdegree C. In the case of laser-induced heating, a LabRAM HR Evolution spectrometer (HORIBA) was used with a 532~nm laser, laser powers between 0.6 and 30~mW, a 100$\times$ objective, a 600~lines/mm grating, and a CCD detector. Spectra were acquired for 30~s with five accumulations. The temperature was determined from the Stokes/anti-Stokes intensity ratio according to: $$\frac{I_{AS}(\omega)}{I_{S}(\omega)} = \left(\frac{\omega_L + \omega}{\omega_L - \omega}\right)^3\exp\left(-\frac{\hbar\omega}{k_BT}\right),$$ where $I_{AS}(\omega)$ and $I_{S}(\omega)$ are the anti-Stokes and Stokes intensities, respectively, $\omega_L = 18787~\mathrm{cm}^{-1}$ ($\lambda = 532.29~\mathrm{nm}$) is the laser frequency, and $T$ is the sample temperature at the laser spot.

Temperature-dependent visible reflectivity spectra were collected using a home-built microspectroscopy setup based on a white LED source (Thorlabs) and an Ocean Optics spectrometer, with a 50$\times$ objective. The reference spectrum was obtained from an aluminum foil prior to measurements on the sample.

\subsection{Data analysis and fitting}
Raman spectra were analyzed by fitting individual phonon modes using Lorentzian line shapes of the form \cite{TDeve07}:
$$I(\omega) = I_0\frac{(\Gamma/2)^2}{(\omega-\omega_0)^2+(\Gamma/2)^2}$$
where $I_0$ is the mode intensity, $\omega_0$ the phonon frequency and $\Gamma$ the full width at half maximum (FWHM).

In cases where electron-phonon coupling leads to an asymmetric line shape, the spectra were fitted using a Fano profile \cite{UFano61}:
$$I(\omega) = \frac{I_0}{q^2}\frac{(q+\epsilon)^2}{(1+\epsilon^2)}, \ \ \mathrm{with} \ \ \ \epsilon = \frac{\omega-\omega_0}{\Gamma/2},$$
where $q$ is the Fano asymmetry parameter, $\omega_0$ the renormalized phonon frequency and $\Gamma$ the linewidth. The inverse magnitude $1/|q|$ is taken as a measure of the electron-phonon coupling strength. A Lorentzian line shape is recovered for $q \rightarrow+\infty$.

The infrared reflectivity spectra were analyzed using the \textsc{RefFIT} software \cite{AKuz05supp} within a multilayer optical model (see examples in Fig.~\ref{fig:figS7}). The diamond anvil was described by a dielectric constant $\varepsilon_d = 5.84$, corresponding to a refractive index $n_d = 2.42$. The complex dielectric function of the sample was modeled using a Drude-Lorentz approach:
$$\varepsilon(\omega) = \varepsilon_{\infty} + \sum_j \frac{\omega_{p,j}^2}{\omega_{0,j}^2 - \omega^2 - i\Gamma_j\omega},$$
where $\varepsilon_\infty$ is the high-frequency dielectric constant, and $\omega_{p,j}$, $\omega_{0,j}$ and $\Gamma_j$ denote the plasma frequency, resonance frequency and linewidth of the $j$-th Lorentz oscillator, respectively. For $\omega_{0,j} = 0$, the oscillator corresponds to a Drude response of free charges.

The reflectivity at the diamond-sample interface is related to the complex dielectric function through the refractive index $\tilde{n}(\omega) = \sqrt{\varepsilon(\omega)}$ as:
$$R(\omega) = \left|\frac{\tilde{n}(\omega) - n_d}{\tilde{n}(\omega) + n_d}\right|^2,$$
where $n_d$ is the refractive index of diamond. The full reflectivity, including the diamond window and multilayer effects, was fitted using the \textsc{RefFIT} software.

To better constrain the high-frequency response, high-temperature visible reflectivity measurements (performed without the diamond window, see Fig.~\ref{fig:figS5}) were included in the fitting procedure.

\FloatBarrier
\bibliography{biblio}

\onecolumngrid

\newpage
\section*{\label{sec:Add}Section 2. Additional data and analysis}

\setcounter{figure}{0}
\renewcommand{\thefigure}{S\arabic{figure}}

\begin{figure*}[h]
\includegraphics[scale=0.4]{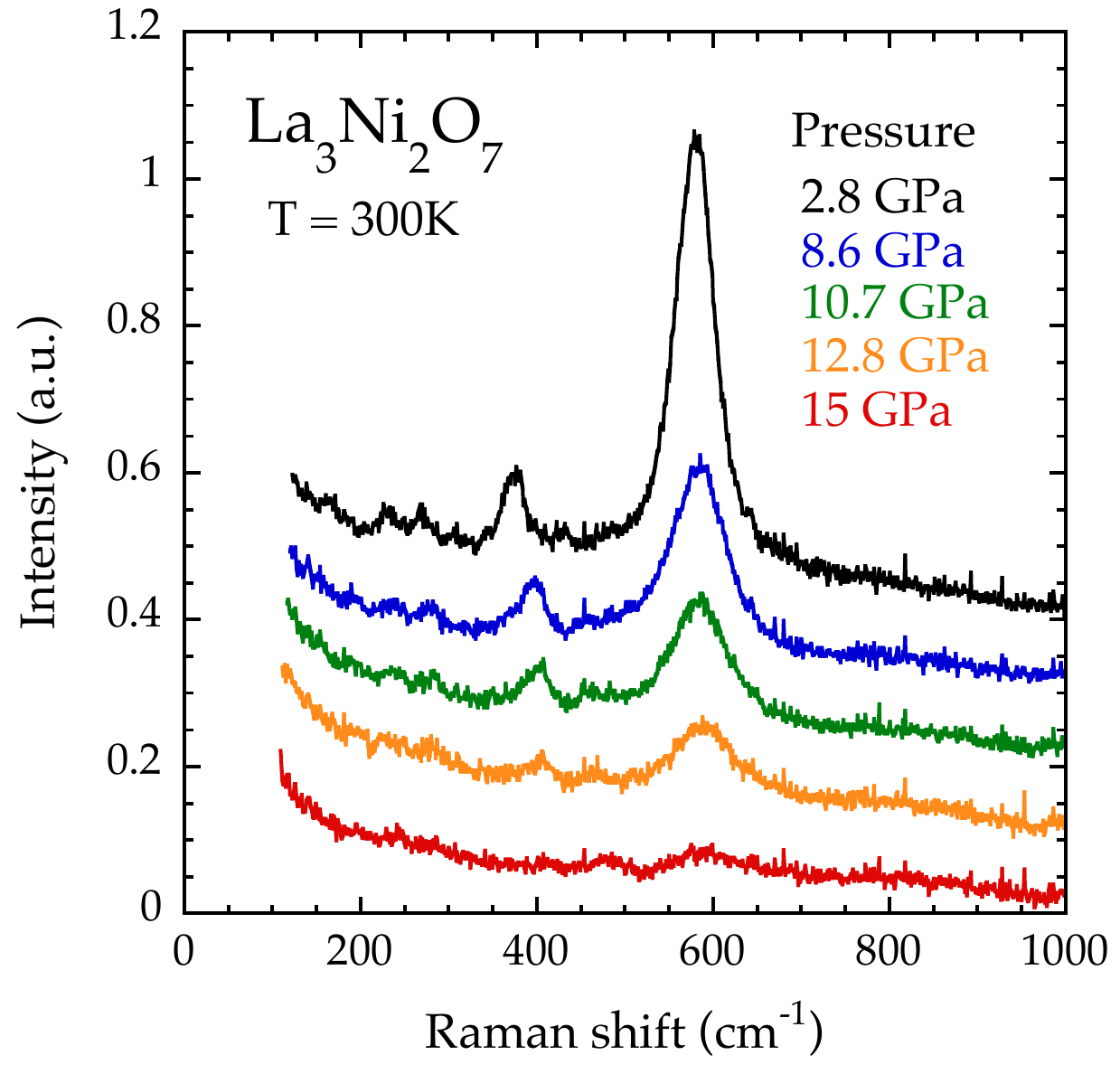}
\caption{\label{fig:figS1} Extra HP Raman data not plotted in Fig.~1 for clarity but fitted and used in Fig.~2.}
\end{figure*}

\begin{figure*}
\centering
\begin{minipage}{.5\textwidth}
  \centering
  \begin{overpic}[scale=0.4]{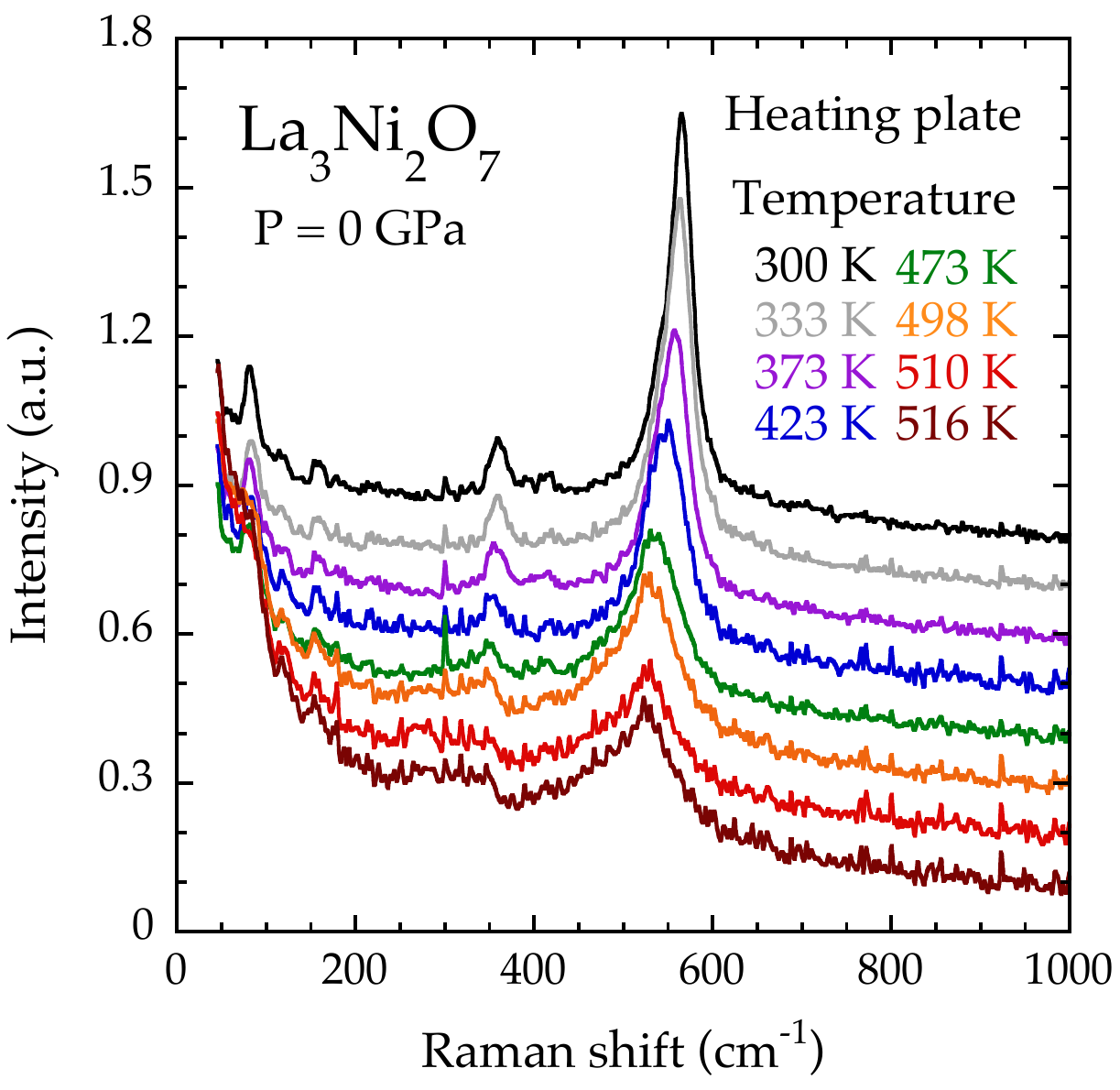}
\put(2,90){\textbf{a)}}
\end{overpic}
\end{minipage}%
\begin{minipage}{.5\textwidth}
  \centering
  \begin{overpic}[scale=0.4]{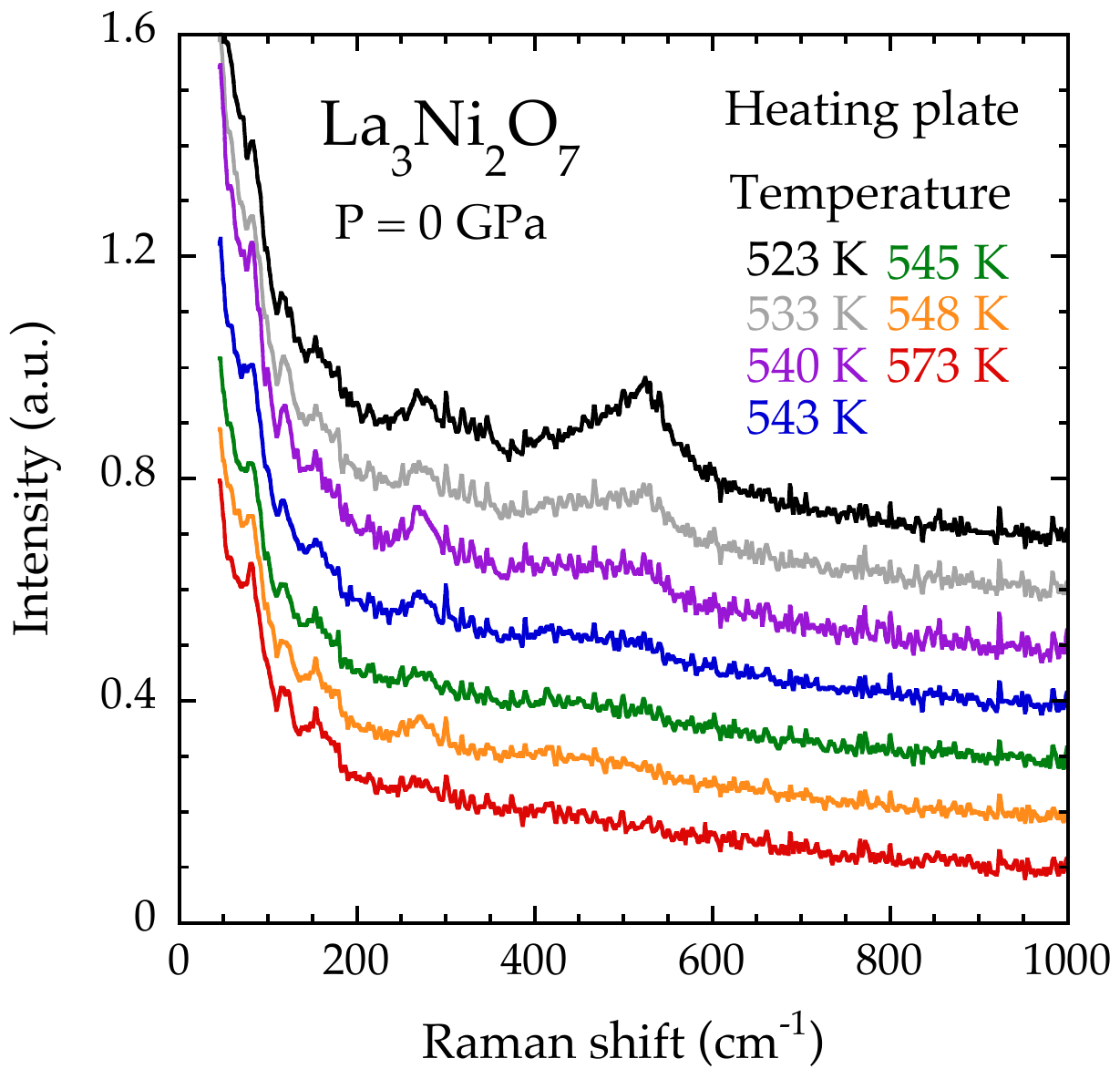}
\put(2,90){\textbf{b)}}
\end{overpic}
\end{minipage}
\caption{\label{fig:figS2} HT Raman data obtained upon heating inside an optical heating stage, complementing the HT data with laser heating in Fig.~1 and the fitting parameters in Fig.~2.}
\end{figure*}

\begin{figure*}
\centering
\begin{minipage}{.5\textwidth}
  \centering
  \begin{overpic}[scale=0.4]{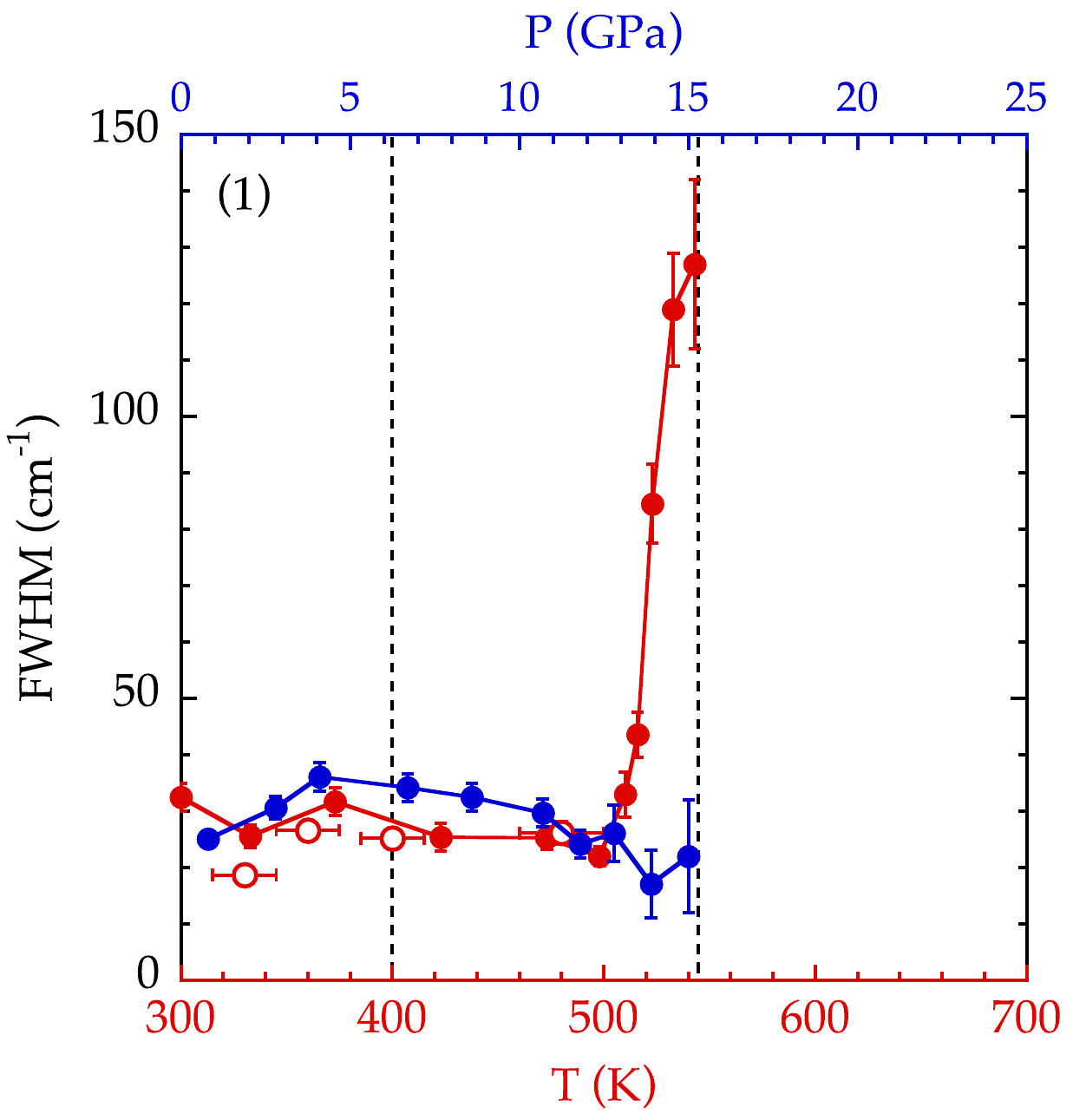}
\put(2,90){\textbf{a)}}
\end{overpic}
\end{minipage}%
\begin{minipage}{.5\textwidth}
  \centering
  \begin{overpic}[scale=0.4]{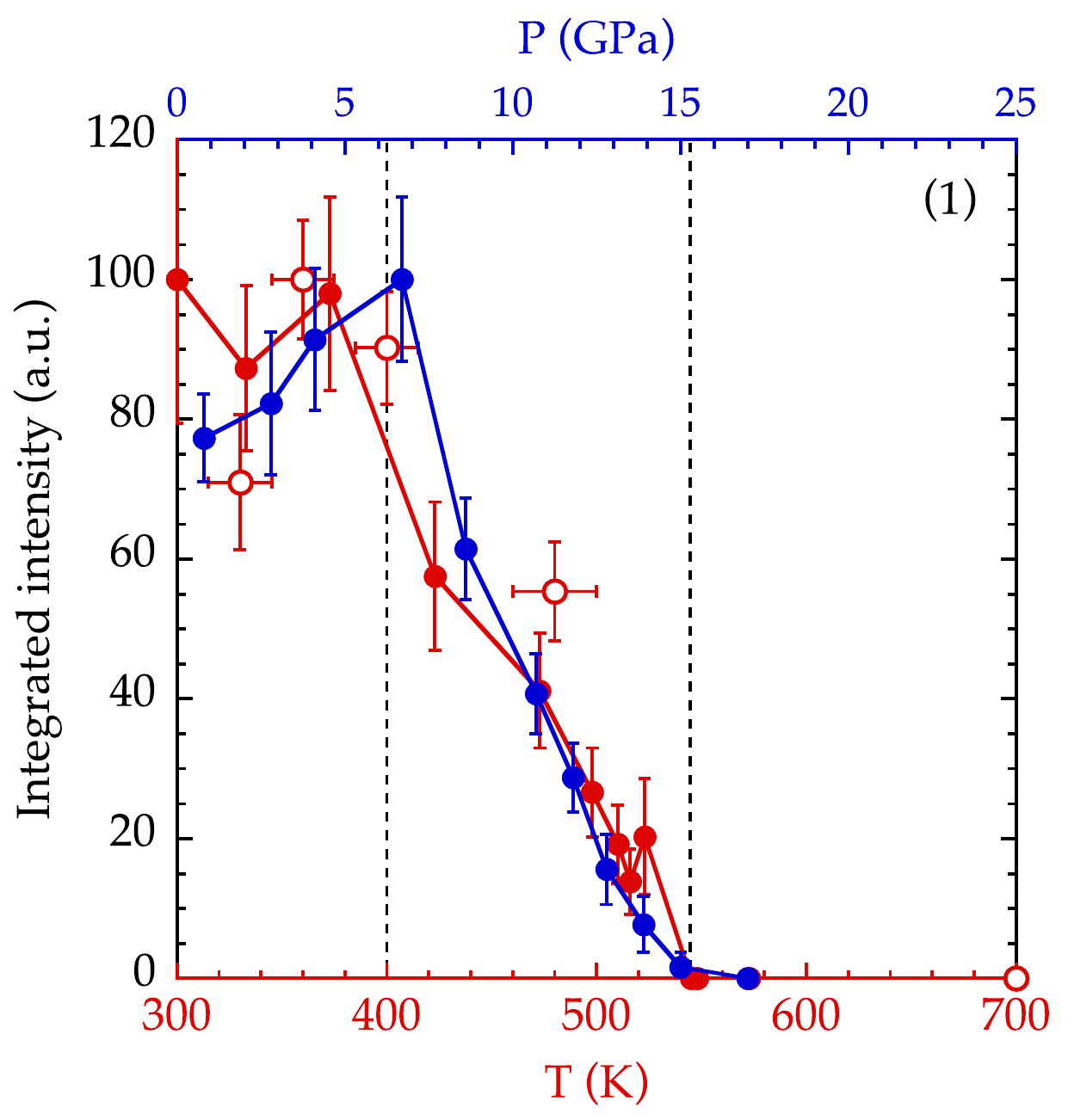}
\put(2,90){\textbf{b)}}
\end{overpic}
\end{minipage}
\caption{\label{fig:figS3} FWHM and integrated intensity of phonon (1) as a function of pressure and temperature obtained from Lorentzian and Fano line-shape fitting, complementing the fitting parameters of phonon (3) in Fig.~2. As phonon (3), this phonon disappears across the structural transition.}
\end{figure*}

\begin{figure*}
\centering
\begin{minipage}{.5\textwidth}
  \centering
  \begin{overpic}[scale=0.4]{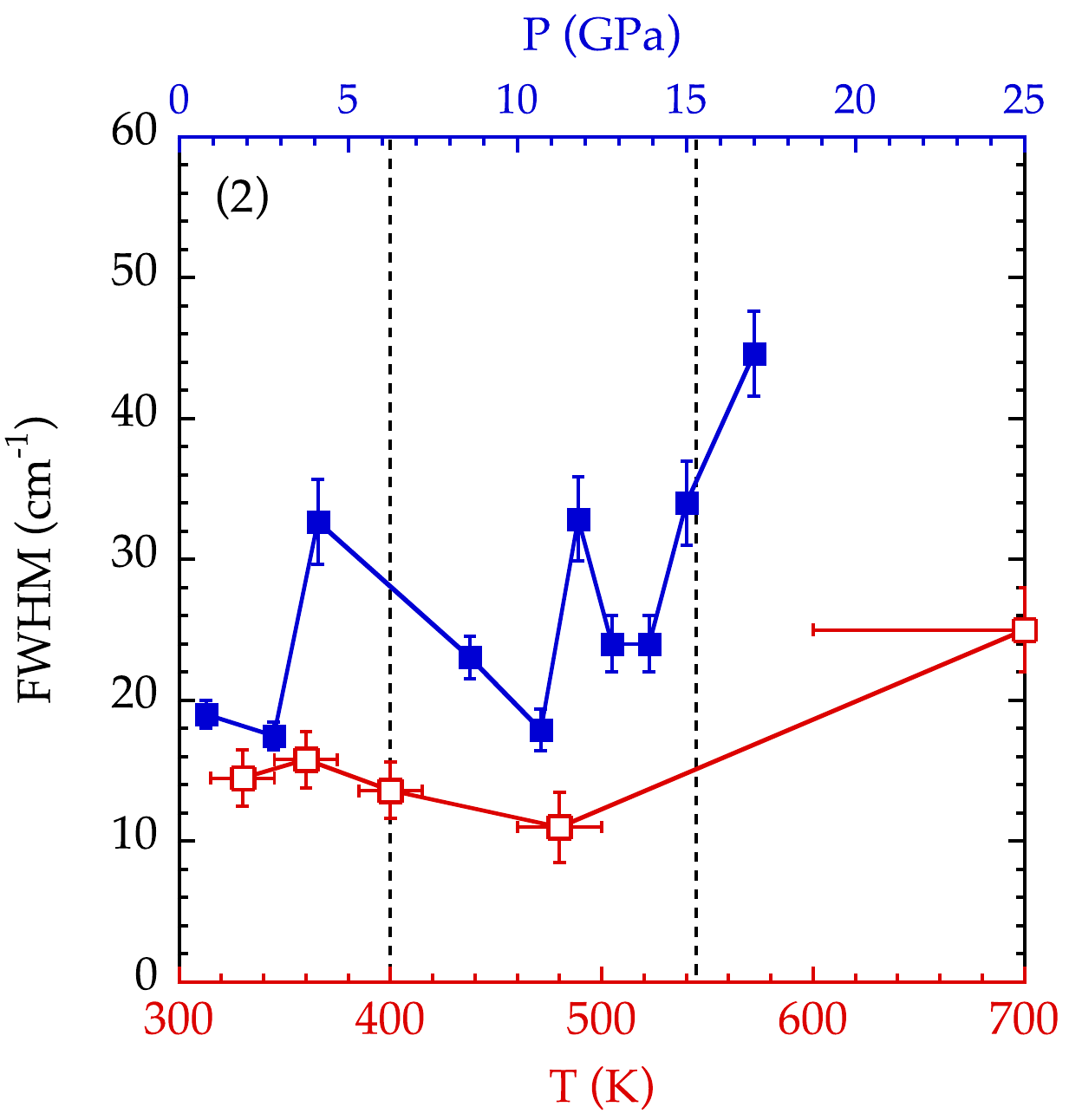}
\put(2,90){\textbf{a)}}
\end{overpic}
\end{minipage}%
\begin{minipage}{.5\textwidth}
  \centering
  \begin{overpic}[scale=0.4]{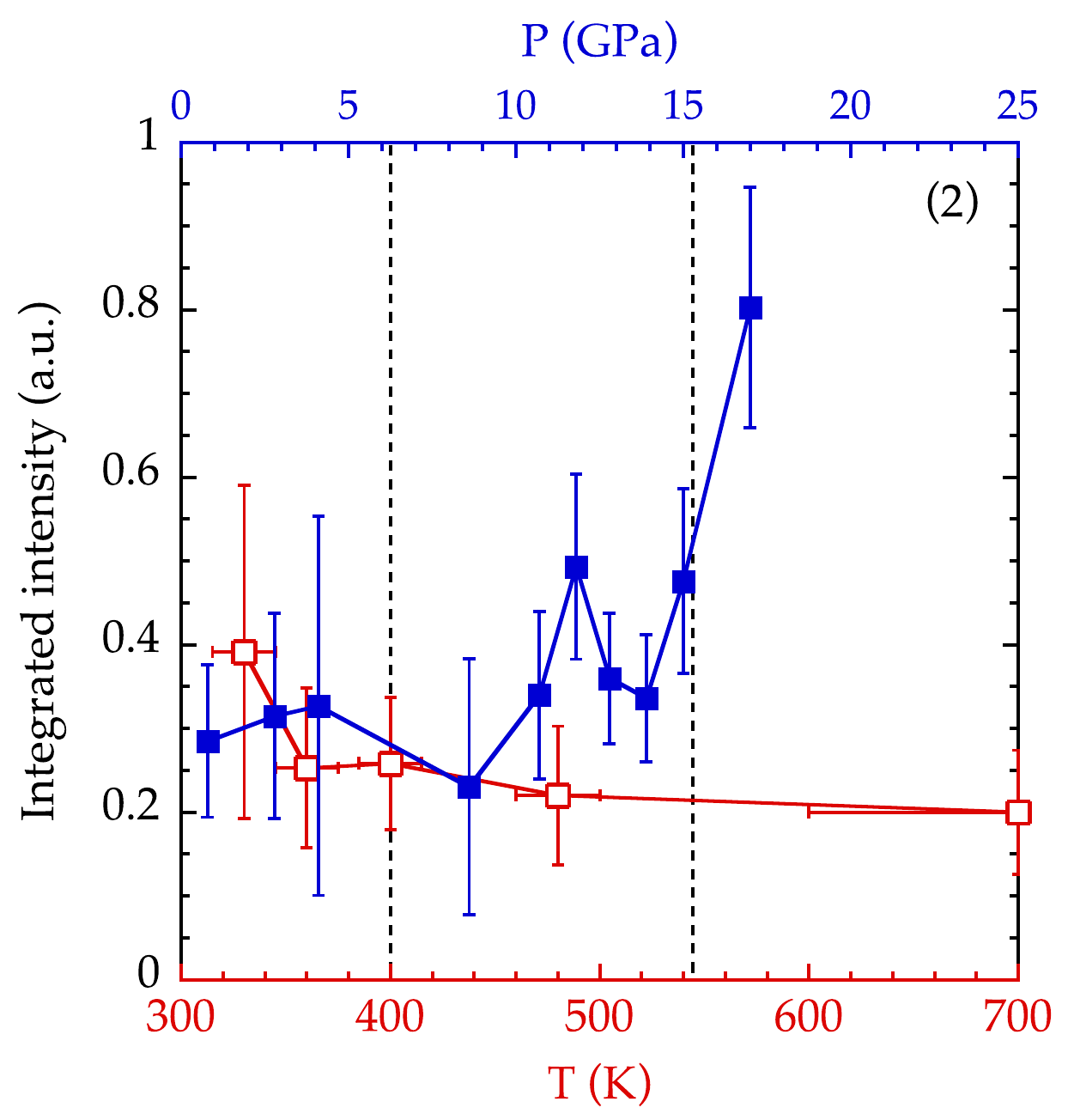}
\put(2,90){\textbf{b)}}
\end{overpic}
\end{minipage}
\caption{\label{fig:figS4} FWHM and integrated intensity of phonon (2) as a function of pressure and temperature obtained from Lorentzian line-shape fitting, complementing the fitting parameters of phonon (3) in Fig.~2. Compared to phonon (1) and (3), this phonon remains across the structural transition.}
\end{figure*}

\begin{figure*}[h]
\includegraphics[scale=0.4]{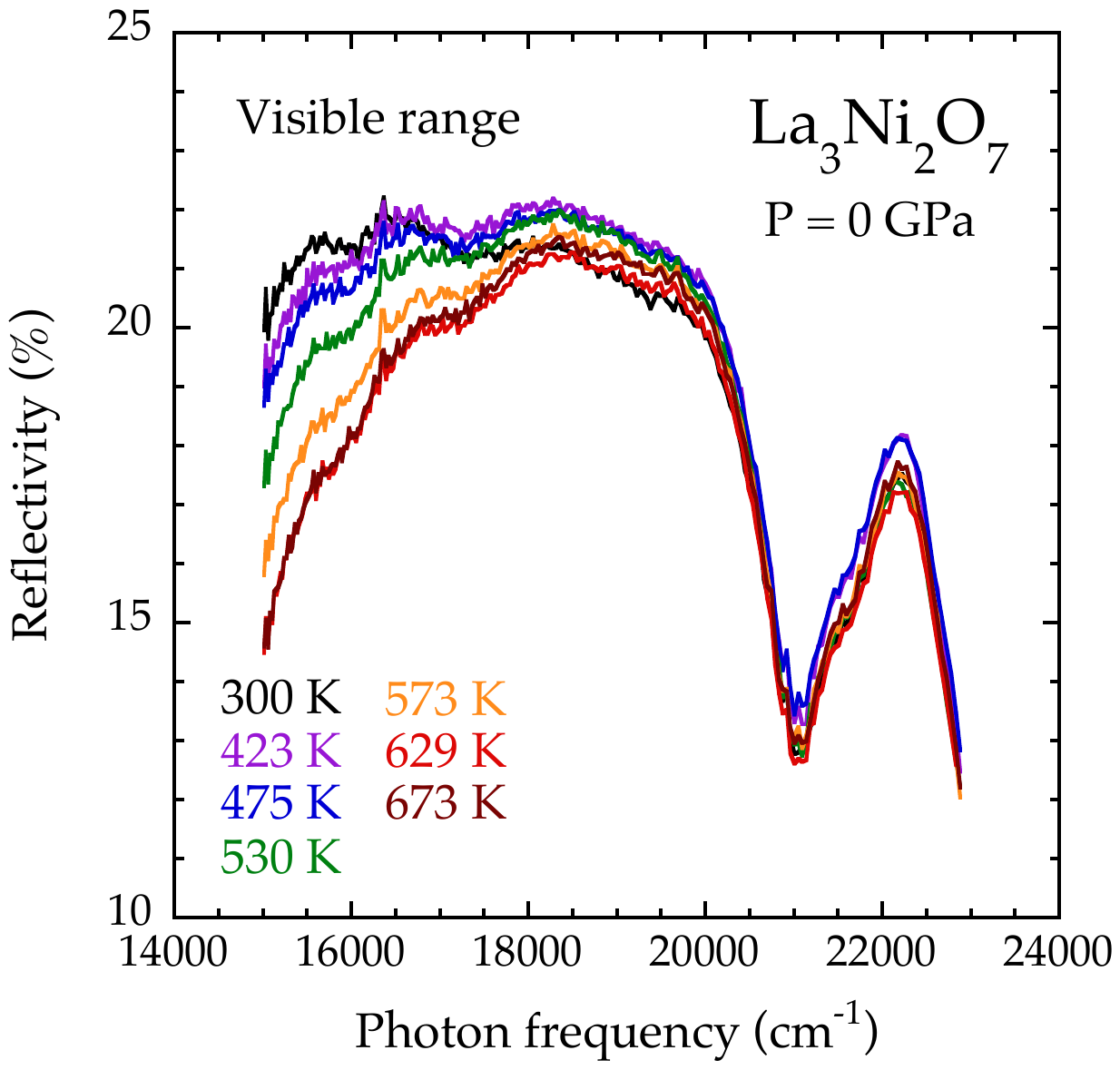}
\caption{\label{fig:figS5} Visible reflectivity taken upon heating with the optical heating stage to describe the color change recorded with the camera and to help fitting the infrared reflectivity data at HP by fixing the high frequency range.}
\end{figure*}

\begin{figure*}
\centering
\begin{minipage}{.5\textwidth}
 \centering
  \begin{overpic}[scale=0.4]{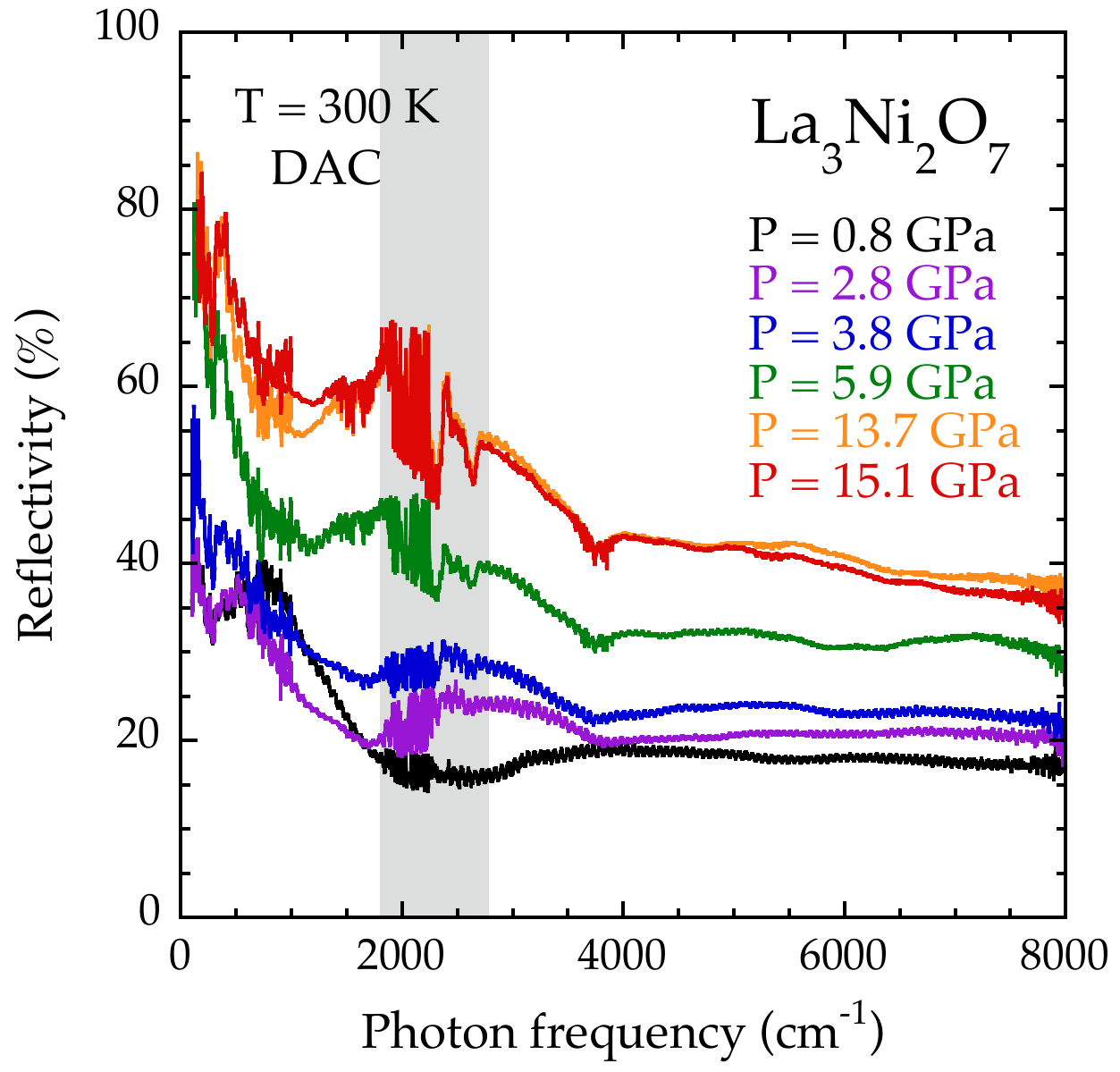}
\put(2,90){\textbf{a)}}
\end{overpic}
\end{minipage}%
\begin{minipage}{.5\textwidth}
  \centering
  \begin{overpic}[scale=0.4]{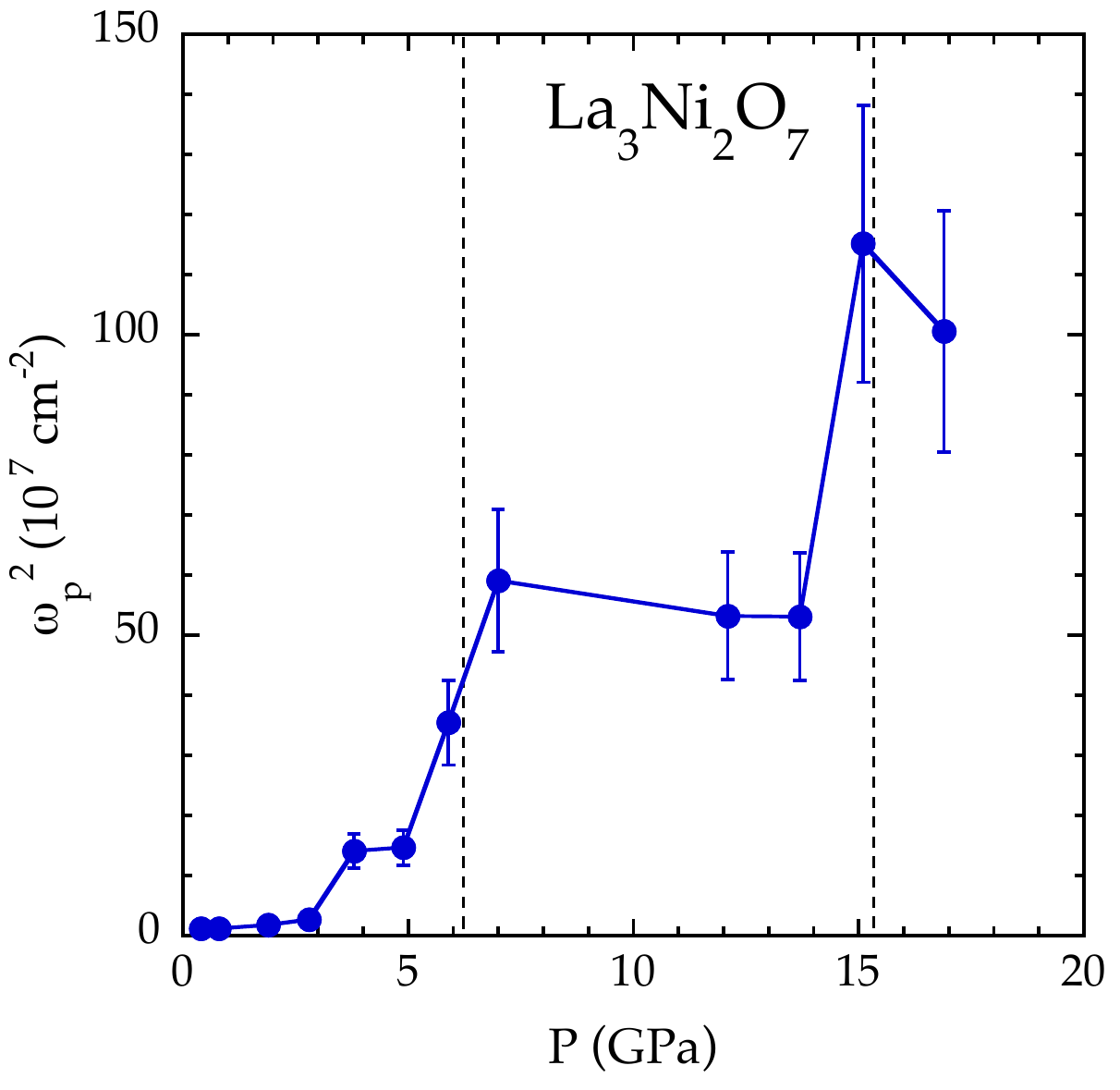}
\put(2,90){\textbf{b)}}
\end{overpic}
\end{minipage}
\caption{\label{fig:figS6} a) Extra HP infrared reflectivity data taken in a diamond anvil cell which are not plotted in Fig.~3a for clarity, but fitted and used in Fig.~3b to extract the plasma frequency $\omega_p$ as a function of pressure. The grayed region corresponds to strong absorption by diamond phonons. -- b) Plasma frequency square $\omega_p^2$ as a function of pressure to illustrate the tremendous enhancement of carrier density through the structural transition.}
\end{figure*}

\begin{figure*}
\centering
\begin{minipage}{.5\textwidth}
 \centering
  \begin{overpic}[scale=0.4]{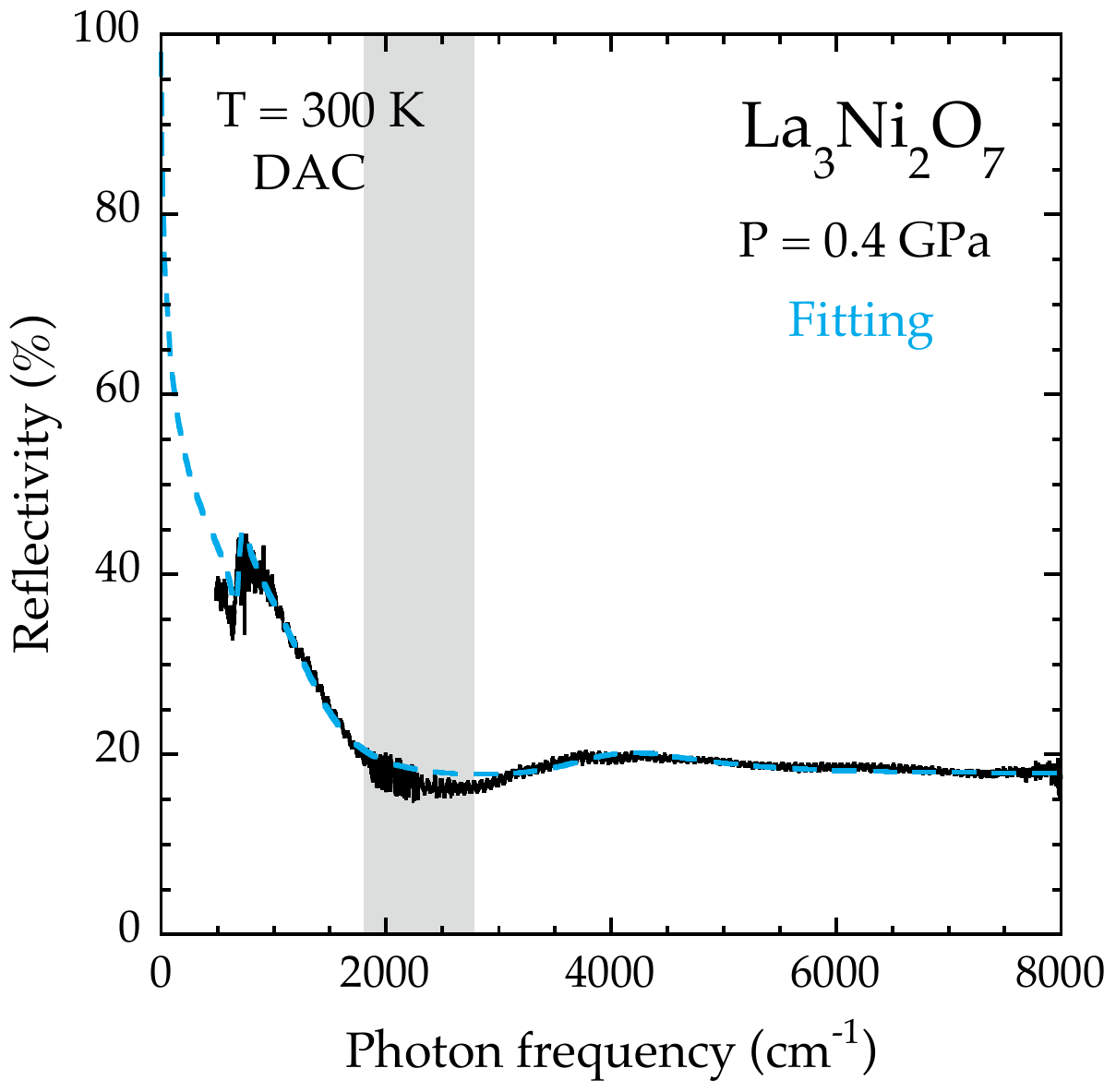}
\put(2,90){\textbf{a)}}
\end{overpic}
\end{minipage}%
\begin{minipage}{.5\textwidth}
  \centering
  \begin{overpic}[scale=0.4]{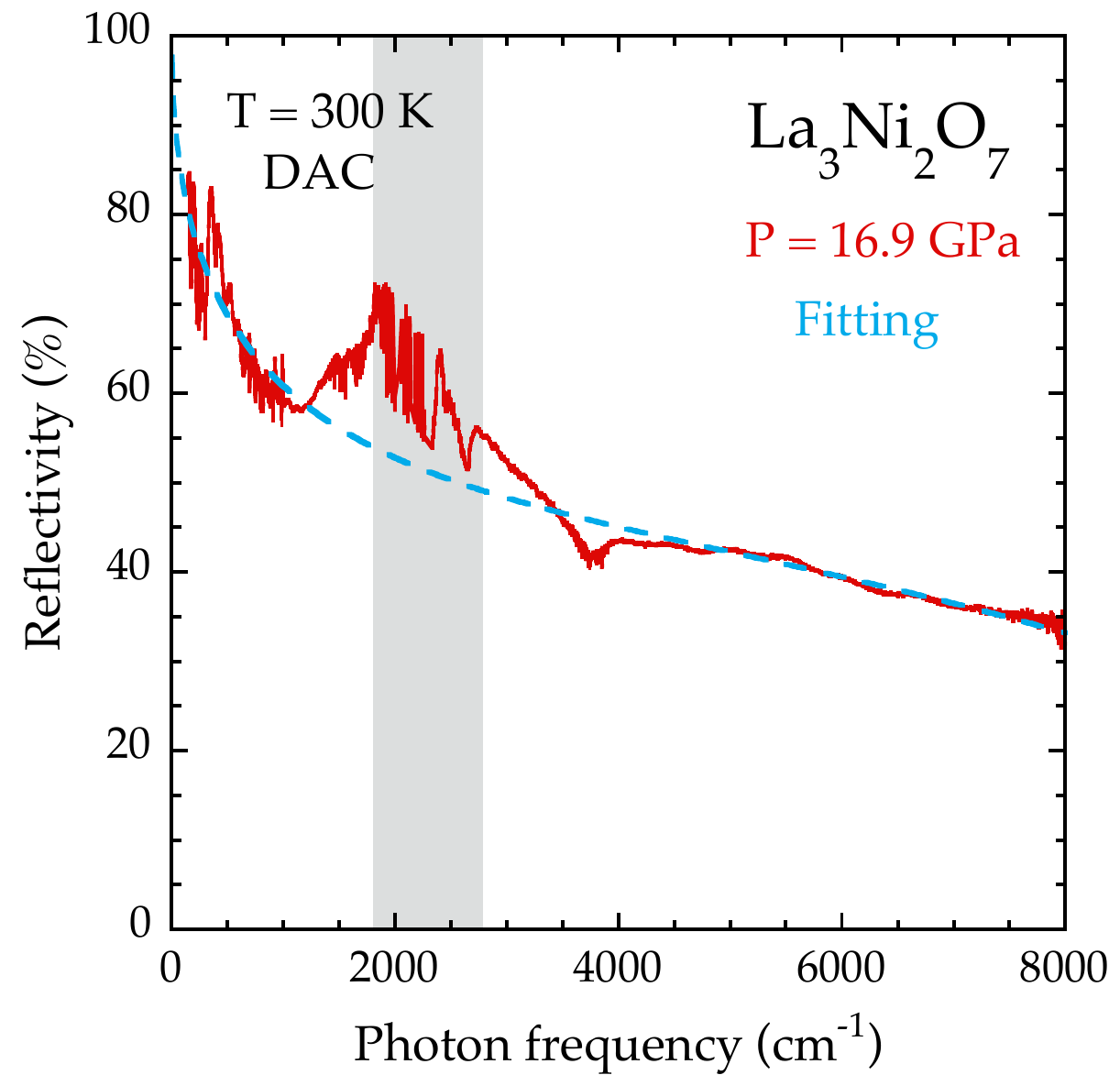}
\put(2,90){\textbf{b)}}
\end{overpic}
\end{minipage}
\caption{\label{fig:figS7} Selected high-pressure infrared reflectivity data and their corresponding multilayer fits in blue dashed lines (Methods) \cite{AKuz05supp}. a) Raw reflectivity spectrum at 0.4~GPa measured through the diamond anvil cell (black curve), together with the \textsc{RefFIT} fit. The gray region indicates strong absorption from diamond phonons. -- b) Same as a) at 16.9~GPa (red curve). At high pressure, the reflectivity data within the shaded spectral range are excluded from the fitting due to significant changes in diamond phonon features, which reduce the reliability of the reference-normalized signal. The fits are used to extract the plasma frequency $\omega_p$ (and $\omega_p^2$) as a function of pressure (Fig.~3b and Fig.~\ref{fig:figS6}b), revealing a strong enhancement of carrier density across the structural transition.}
\end{figure*}